\documentclass[onecolumn,10pt]{asme2ej}

\usepackage{cite}
\usepackage{amsmath,amssymb,amsfonts}
\usepackage{algorithmic}
\usepackage{graphicx}
\usepackage{textcomp}
\usepackage{makecell}
\usepackage{amsmath}
\usepackage{multicol, blindtext}
\usepackage{multirow}
\usepackage{tabularx}
\usepackage{csquotes}
\usepackage{xcolor}
\usepackage{float}
\usepackage{bm}
\usepackage{etoolbox}
\usepackage{gensymb}
\usepackage{threeparttable}

\newcolumntype{C}[1]{>{\centering\arraybackslash}p{#1}}

\graphicspath{ {./figure/} }

\usepackage{nomencl}
\makenomenclature

\usepackage{etoolbox}
\renewcommand\nomgroup[1]{%
  \item[\bfseries
  \ifstrequal{#1}{A}{}{%
  \ifstrequal{#1}{B}{Number Sets}{%
  \ifstrequal{#1}{C}{Other Symbols}{}}}%
]}
\renewcommand{\nompreamble}{\begin{multicols}{2}}
\renewcommand{\nompostamble}{\end{multicols}}

\title{Parameter Identification and Sensitivity Analysis for Zero-dimensional Physics-based Lithium-Sulfur Battery Models}

\author{Chu Xu
    \affiliation{
	Mechanical Engineering\\
	University of Maryland\\
    Email: chuxu88@umd.edu
    }	
}

\author{Timothy Cleary
    \affiliation{
	Mechanical Engineering\\
	Penn State University\\
    Email: tdc142@psu.edu
    }	
}

\author{Guoxing Li
    \affiliation{
    Professor at Institute of Frontier and\\
    Interdisciplinary Science\\
    Shandong University\\
    Email: gxli@sdu.edu.cn
    }	
}

\author{Donghai Wang
    \affiliation{
	Professor of Mechanical Engineering\\
	Penn State University\\
    Email: dwang@psu.edu
    }	
}

\author{Hosam Fathy\thanks{Address all correspondence to this author.} \\
    \affiliation{
	Professor, Fellow of ASME\\
	Mechanical Engineering\\
	University of Maryland\\
    Email: hfathy@umd.edu
    }
}

\begin{document}
\maketitle

\begin{abstract}
{\it This paper examines parameter estimation for Lithium-Sulfur (Li-S) battery models from experimental data. Li-S batteries are attractive compared to traditional Lithium-ion batteries, thanks largely to their potential to achieve higher energy densities. The literature presents a number of Li-S battery models with varying fidelity and complexity levels. This includes both high-fidelity diffusion-reaction models as well as zero-dimensional models that neglect diffusion dynamics while capturing the underlying reduction-oxidation reaction physics. This paper focuses on four zero-dimensional models, representing different possible sets of redox reactions. There is a growing need for using experimental data sets to both parameterize and compare these models. To address this, Li-S coin cells were fabricated and tested. In parallel, a sensitivity analysis of key model parameters was conducted. Using this analysis, a subset of model parameters was selected for identification and estimation in all four Li-S battery models. 
}
\end{abstract}


\mbox{}
\nomenclature[A, 01]{$i$}{Species index}
\nomenclature[A, 02]{$j$}{Reaction index}
\nomenclature[A, 03]{$q$}{Number of species}
\nomenclature[A, 04]{$p$}{Number of reactions}
\nomenclature[A, 05]{$m_i$}{Mass of species $i$  [g]}
\nomenclature[A, 06]{$m_{S_p}$}{Mass of precipitated sulfur [g]}
\nomenclature[A, 07]{$i_j$}{Current generated by reaction $j$  [A]}
\nomenclature[A, 08]{$I$}{Input (discharge) current   [A]}
\nomenclature[A, 09]{$E_j$}{Reduction potential for reaction $j$  [V]}
\nomenclature[A, 10]{$\eta_j$}{Overpotential for reaction $j$  [V]}
\nomenclature[A, 11]{$V$}{Output voltage across the cell  [V]}
\nomenclature[A, 12]{$\varepsilon$}{Relative porosity}
\nomenclature[A, 13]{$a_v$}{Active reaction area  [m$^2$]}
\nomenclature[A, 14]{$n_{S_i}$}{Number of sulfur atoms in species $i$}
\nomenclature[A, 15]{$n_{j}$}{Number of electrons exchanged in reaction $j$}
\nomenclature[A, 16]{$M_{S}$}{Molar mass of sulfur  [g/mol]}
\nomenclature[A, 17]{$F$}{Faraday's constant  [C/mol]}
\nomenclature[A, 18]{$R$}{Gas constant  [J/(K~mol)]}
\nomenclature[A, 19]{$T$}{Temperature  [K]}
\nomenclature[A, 20]{$v$}{Cell volume  [L]}
\nomenclature[A, 21]{$S_{sat}$}{Saturation mass of $S^{2-}$  [g]}
\nomenclature[A, 22]{$s_{i,j}$}{Stoichiometric coefficients}
\nomenclature[A, 23]{$m_i^0$}{Initial mass of species $i$  [g]}
\nomenclature[A, 24]{$E_j^0$}{Standard reduction potential  [V]}
\nomenclature[A, 25]{$i_j^0$}{Exchange current density  [$A/m^2$]}
\nomenclature[A, 26]{$a_v^0$}{Initial active reaction area  [m$^2$]}
\nomenclature[A, 27]{$\gamma$}{Power of the relative porosity}
\nomenclature[A, 28]{$\omega$}{Relative porosity change rate constant [1/g]}
\nomenclature[A, 29]{$k_p$}{Precipitation rate constant [1/(g~s)]}
\printnomenclature

\section{Introduction}

As the performance of Lithium-ion (Li-ion) batteries approaches its practical limit (up to 500 Wh/kg) \cite{wild2019lithium}, researchers are actively seeking alternative, high energy density solutions. The Lithium-Sulfur (Li-S) chemistry is one such solution. It has attracted significant attention from the scientific community for at least four reasons. First,  the Li-S chemistry offers very attractive theoretical limits for specific capacity (1672 $Ah/kg$) and specific energy (2600 $Wh/kg$) \cite{wild2019lithium}. Prototype Li-S cells have already achieved specific energies around 700 Wh/kg~\cite{yanin1983low}. Second, Li-S batteries have the potential to offer a range of operating temperatures from -40 to 60 \degree C \cite{huang2013entrapment}. This is especially beneficial for low temperature applications. Third, phenomena such as precipitation and the internal charge shuttle effect provide Li-S batteries with some degree of intrinsic overcharge protection~\cite{hunt2015lithium, mikhaylik2004polysulfide}, thereby making them less vulnerable to catastrophic failure compared to their Li-ion counterparts. Fourth, the use of sulfur, instead of rare earth materials, makes Li-S batteries quite appealing in terms of production cost and environmental footprint \cite{bresser2013recent}. In spite of these benefits, the Li-S chemistry suffers from low cycle life and relatively high self-discharge rates \cite{manthiram2014rechargeable,song2013lithium}. Efforts are ongoing to improve the electrochemical performance of Li-S batteries \cite{rezan2017li}. This paper is part of a complementary effort to improve Li-S battery performance through modeling, estimation, and, ultimately, optimal control. The paper focuses specifically on the problem of parameterizing Li-S battery models from experimental data: a critical first step towards model-based control. 

The literature already presents a number of Li-S battery models that can be used for parameter estimation and model-based control. These models fall on a spectrum of fidelity and complexity levels. Equivalent circuit models (ECMs) of Li-S batteries sit at one end of this spectrum. They typically combine empirical open-circuit voltage maps with (often charge- and temperature-dependent)  resistor-capacitor circuits~\cite{kolosnitsyn2011study, deng2013electrochemical, knap2015electrical,propp2016multi}. Typical uses of such simple models include online state of charge estimation~\cite{fotouhi2017lithium, fotouhi2017lithium_sul,propp2019improved}. At the other end of the spectrum, one finds physics-based models of the reduction-oxidation, diffusion, precipitation, and self-discharge dynamics in Li-S batteries. These are typically partial differential algebraic equation (PDAE) models of coupled diffusion-reaction dynamics~\cite{kumaresan2008mathematical, fronczek2013insight, hofmann2014mechanistic, ren2016modeling, andrei2018theoretical,danner2019influence}. They have the advantage of providing higher-fidelity representations of the underlying battery physics, at the expense of high complexity. Zero-dimensional (0D) models provide an important middle ground between these two modeling extremes. They neglect the diffusion of species inside Li-S batteries, but use the laws of electrochemistry to model the underlying battery reactions. This approach results in lower-order models with fewer parameters to identify and reduced computational complexity. Simplified two-reaction and multi-reaction 0D models introduced in \cite{marinescu2016zero} and \cite{zhang2015modeling} are able to capture the main features of battery voltage response during discharge. The computational complexity of a 0D Li-S battery model depends on the specific choice of which reactions to model, and which to ignore. One potential approach for making this choice is to compare the fidelity with which different 0D models can fit a given experimental data set. Such an exercise is appealing because it can furnish both: (i) a systematic comparison between different battery models using the same data set, and (ii) an experimental parameterization of different 0D Li-S battery models. To the best of the authors' knowledge, both of these contributions remain relatively unexplored in the literature. The overarching goal of this work is to address this research gap. 

Starting from previous work on 0D Li-S models, this work goes one step further to analyze different 0D models capturing different underlying redox reactions. The goal is to answer the following questions: (i) How does each reaction influence the discharge voltage curve in a 0D model? (ii) Do all the model parameters have a significant impact on the discharge voltage curve? Is it necessary to identify them all? (iii) How accurately can the identified models fit the experimental voltage behavior? Is there a model with a specific reaction chain that performs best? The ultimate goal is to provide guidance on the choice of which reactions to model for future applications, such as state of charge estimation. 

The remainder of this paper addresses these questions, and is organized as follows. Section 2 describes the 0D model structure adopted from \cite{zhang2015modeling,marinescu2016zero}, combined with the reduction reactions considered in each model. Section 3 provides a simulation study illustrating the effects of different reactions on the discharge voltage performance. Section 4 performs a parameter sensitivity study to provide guidance on which parameters are critical to identify from experimental data. Section 5 presents the parameter identification results using experimental data. Finally, Section 6 summarizes this paper's conclusions. These conclusions are based on an experimental study at a discharge rate of 0.3C. At this discharge rate, the discharge voltage curve of an Li-S battery exhibits three main features, namely: a high plateau, a low plateau, and a dip point separating these two plateaus. In the high plateau region, the active material, $S_8$, in the cathode accepts electrons to produce the polysulfide, $S_x^{2-}$ ($x$ can be 8, 6, 4). Further polysulfide reduction takes place in the lower voltage plateau region \cite{wild2015lithium}. Figure 1 shows two discharge characteristic curves, obtained by cycling Li-S cells fabricated as part of this study at 0.25C and 0.125C. Both of these curves exhibit the three main features discussed above, namely: the high plateau, low plateau, and dip point. Moreover, the small differences in discharge voltage between these curves can be partially attributed to simple polarization effect, such as Ohmic polarization. Thus, the paper’s focus on a discharge C-rate of 0.3C can help shed a light on behaviors observed over a range of C-rates, especially C-rates below 0.3C. 
 
\begin{figure}[ht]
    \centering
    \includegraphics[ width=0.4\textwidth]{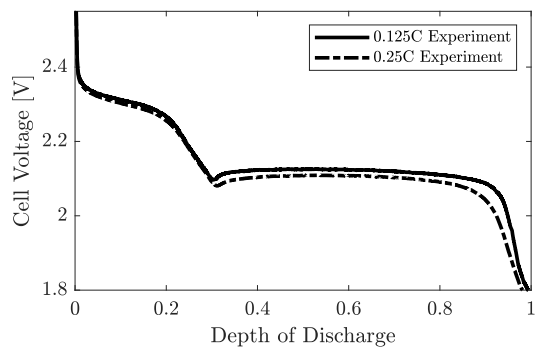}
    \caption{Experimental cell voltage profiles under different discharge currents}
    \label{fig:Discharge_Crates}
\end{figure}

\section{Zero-dimensional Physics-based Model}
This section describes the 0D reaction models used for subsequent analysis. The models build on two existing bodies of work. First, research by Zhang \textit{et al.} \cite{zhang2015modeling} develops a lumped model describing the dynamics of dissolved species concentrations and cathode material porosity in Li-S batteries. Moreover, this model includes the full reduction reaction chain and the effective electrolyte resistance. Second, research by Marinescu \textit{et al.} \cite{marinescu2016zero} introduces another 0D model taking into account the dynamic evolution of different species' masses as well as the shuttle effect. This model simplifies the chain of sulfur reduction reactions into two steps. Our work combines the underlying structures of these two models, and expands them to capture the dynamics of different sets of reduction reactions. The end result is a set of four 0D models ranging from low to high complexity. The simplest model covers a minimum set of reduction reactions modeled in the literature, whereas the most complex model covers the full set of sulfur reduction reactions in Li-S batteries. The reactions considered in each model are listed in Table \ref{tab:Reactions}. In this work, the authors adopt the following modeling assumptions from the literature: (i) there is an unlimited amount of lithium supply in the cell with a negligible overpotential on the anode side, as in \cite{fronczek2013insight}; (ii) the shuttle effect of polysulfides is not included due to this paper's focus on the voltage performance instead of capacity fade \cite{zhang2015modeling}; and (iii) only the lowest polysulfide's precipitation reaction ($2Li^{+}+S^{2-} \rightleftharpoons Li_2S \downarrow$) is modeled \cite{zhang2015modeling,marinescu2016zero}.  

\begin{table*}[ht]
	\begin{center}
		\caption{Reduction reactions considered in the models}
		\label{tab:Reactions}
		\begin{tabular}{|C{2.2cm}|C{2.8cm}|C{2.8cm}|C{3.3cm}|C{3.5cm}|}
			\hline 
			\textbf{Models} & \textbf{Model \#1} & \textbf{Model \#2} &  \textbf{Model \#3}  &   \textbf{Model \#4} \\ 
			\hline 
			\multirow{5}{*}{ \makecell{\textbf{Reduction}\\\textbf{Reactions}}}  &
			$ \frac{1}{4} S_8 \text{~~} +e^- \rightleftharpoons \frac{1}{2} S_4^{2-}  $ 
			& $ \frac{3}{8} S_8 \text{~~} +e^- \rightleftharpoons \frac{1}{2} S_6^{2-}  $
			& $ \frac{1}{2} S_8 \text{~~} +e^- \rightleftharpoons \frac{1}{2}  S_8^{2-}  $
			& $ \frac{1}{2} S_8 \text{~~} +e^- \rightleftharpoons \frac{1}{2}  S_8^{2-}  $ \\

			&
			$ \frac{1}{6} S_4^{2-} +e^- \rightleftharpoons \frac{2}{3}  S^{2-}  $
			& $  \text{~} S_6^{2-}  +e^- \rightleftharpoons \frac{3}{2} S_4^{2-}  $
			& $ \frac{3}{2} S_8^{2-}  +e^- \rightleftharpoons 2 S_6^{2-}  $
			& $ \frac{3}{2} S_8^{2-}  +e^- \rightleftharpoons 2 S_6^{2-}  $ \\

		    &
			& $ \frac{1}{6} S_4^{2-} +e^- \rightleftharpoons \frac{2}{3}  S^{2-}  $
			& $  \text{~~} S_6^{2-}  +e^- \rightleftharpoons \frac{3}{2} S_4^{2-}  $
			& $ \text{~~} S_6^{2-}  +e^- \rightleftharpoons \frac{3}{2} S_4^{2-}  $ \\

		    &
			& 
			& $ \frac{1}{6} S_4^{2-} +e^- \rightleftharpoons \frac{2}{3}  S^{2-}  $
			& $ \frac{1}{2} S_4^{2-} +e^- \rightleftharpoons  \text{~~}  S_2^{2-}  $ \\

		    &
		   	& 
			& 
			& $ \frac{1}{2} S_2^{2-} +e^- \rightleftharpoons   \text{~~} S^{2-}  $ \\
			\hline
		 \makecell{\textbf{Dissolved} \\ \textbf{Sulfur Species}} & $S_8,~S_4^{2-},S^{2-}$ & $S_8,~S_6^{2-},S_4^{2-},S^{2-}$ & $S_8,~S_8^{2-}, S_6^{2-},S_4^{2-},S^{2-}$ & $S_8,~S_8^{2-}, S_6^{2-},S_4^{2-},S_2^{2-},S^{2-}$ \\	
		\hline
		\end{tabular}
	\end{center}
\end{table*}

Since the model is zero-dimensional, there exists no mass transport due to diffusion/migration. The only dynamics changing the masses of the dissolved species are the electrochemical reaction and precipitation. These dynamics lead to the time-evolution relations for the masses of various sulfur species and the porosity of the cathode material, which act as the state variables in the model. Below is the resulting differential algebraic equation (DAE) model (assuming discharge), including both state equations and algebraic constraints. Descriptions of the model's variables and parameters are included in the paper's nomenclature section. \\
\textbf{State Equations:}
\begin{align}
    \dot{m}_i &= \sum_j \frac{n_{S_i} M_s}{n_jF}s_{i,j}i_j  \label{mass1} ~, 
    \text{~~~for~}i = 1,...,q-1 \\
    \dot{m}_{q} &= \frac{n_{S_q} M_s}{n_p F}s_{q,p}i_p - \dot{m}_{S_p} \label{mass2} \\
    \dot{m}_{S_p} &= k_p m_{S_p} ({m}_{q}-S_{sat}) \label{mass3} \\
    \dot{\varepsilon} &= - \omega \dot{m}_{S_p}  \label{poro}
\end{align}
\textbf{Constraints:}
\begin{align}
    E_j =   E_j^0  - &  \frac{RT}{n_j F} \sum_i s_{i,j}\text{~ln}(\frac{m_i}{n_{S_i} M_s v}) 
    \label{E_j}\\
    i_j = -a_v i_j^0 \{\prod_i ( \frac{m_i}{m_i^0})^{s_{i,j}}  &  e^{\frac{F}{2RT}\eta_j} 
    - \prod_i (\frac{m_i}{m_i^0})^{-s_{i,j}} e^{-\frac{F}{2RT}\eta_j}\} \label{ij}\\
    a_v &= a_v^0 \varepsilon^\gamma \label{activeArea}\\
    I &= \sum_j i_j \label{currentsum} \\
    V &= \eta_j + E_j \label{VoltageEqn}
\end{align}

The rates of mass change for higher-order dissolved sulfide species are governed by Eqn. \ref{mass1}. For species $S^{2-}$, one needs to consider its mass generation from the last reduction reaction and its mass loss due to precipitation, as shown in Eqn. \ref{mass2}. The nucleation and growth phenomenon is described by Eqn. \ref{mass3} \cite{marinescu2016zero}. The rate of change of precipitate mass is partly driven by the precipitate mass itself. This reflects the fact that the existing precipitate serves as a nucleus for further precipitation/growth, as long as the mass in the electrolyte is above a given saturation mass. 

Relative porosity is one of the state variables in the model with the dynamics governed by Eqn. \ref{poro}. This variable equals the current porosity of the cathode material divided by initial porosity, and has a direct effect on the active reaction area, as shown in Eqn. \ref{activeArea}  \cite{kumaresan2008mathematical}. When the porosity decreases to zero, all the reactants are blocked by the precipitate from the cathode material surface. This can be one of the indicators of the cell's full discharge, another full discharge scenario being the reduction of all polysulfides to $S^{2-}$. As the cell approaches complete coverage of the cathode ($\varepsilon \rightarrow 0$), larger overpotentials ($\eta_j$) are needed for the same current. This provides one explanation of the low measured voltage at the end of Li-S battery discharge. 

The reduction potential of each reaction is given by the Nerst equation (Eqn. \ref{E_j}), assuming that $E_j^0$ is the equilibrium potential when the concentration of the dissolved species in reaction $j$ is 1 mol/L \cite{zhang2015modeling}. The current generated by the corresponding reduction reaction is described by the Butler–Volmer equation (Eqn. \ref{ij}). All these currents sum to the only input variable, namely, the external discharge current $I$, as shown in Eqn. \ref{currentsum}. Finally, Eqn. \ref{VoltageEqn} states the relation between the voltage measured across the battery, the overpotential and the reduction potential of each reaction.

\section{Voltage Responses for Different Reaction Chains}

\begin{figure}[ht]
    \centering
    \includegraphics[width=0.5\textwidth]{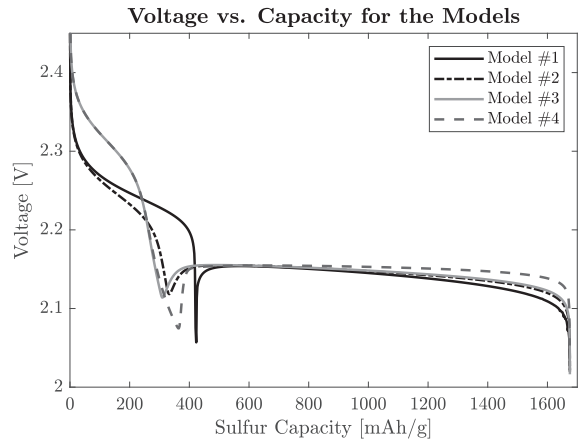}
    \caption{Discharge voltages vs. sulfur capacity for the reaction models}
    \label{fig:reactionCompare}
\end{figure}

In the literature, there exist at least two simplifications of the reaction chain on the cathode side, namely, the two-step reduction $S_8 \rightarrow S_4^{2-} \rightarrow S^{2-}$ \cite{danner2015modeling} and the four-step reduction $S_8 \rightarrow S_8^{2-} \rightarrow S_6^{2-}\rightarrow S_4^{2-} \rightarrow S^{2-}$ \cite{ren2016modeling}. Different Li-S battery models, capturing different representations of the overall reaction chain, are built to suit the focus areas of the corresponding research studies. In this section, the authors compare the simulated voltage outputs of the models in Table \ref{tab:Reactions} under a 0.3C discharge, and analyze each reaction's influence on the different features of the discharge voltage curve. OpenModelica \cite{fritzson2006openmodelica} is used to build and simulate the models with the key parameter values listed in Table \ref{tab:para_value_fix}. The simulated discharge voltage-capacity curves are compared in Fig. \ref{fig:reactionCompare}.

\begin{table}[ht]
\caption{Nominal parameter values in the simulations}
\label{tab:para_value_fix}
\centering
  \begin{threeparttable}
		\begin{tabular}{|C{2.2cm}|C{1.8cm}|C{4.5cm}|C{1.5cm}|}
			\hline 
			\textbf{Model} & \textbf{Notations} & \textbf{~Values} & \textbf{Units}  \\ 
			\hline 
			\multirow{2}{*}{\textbf{Model \#1}} & 
			$E_j^0 ~^*$ & 2.40,~~2.10 & $V$ \\ 
			&$i_j^0~^{**}$ & 2.00,~~0.02 & $A/m^2$ \\  \hline
			\multirow{2}{*}{\textbf{Model \#2}} &   
	    	$E_j^0 ~^*$ & 2.40,~~2.30,~~2.10 & $V$ \\
			&$i_j^0~^{**}$ & 2.00,~~0.02,~~0.02 & $A/m^2$ \\  \hline
			\multirow{2}{*}{\textbf{Model \#3}} &  
			$E_j^0 ~^*$ & 2.46,~~2.38,~~2.30,~~2.10 & $V$ \\
			&$i_j^0~^{**}$ & 2.00,~~0.02,~~0.02,~~0.02 & $A/m^2$ \\ \hline
			\multirow{2}{*}{\textbf{Model \#4}} &  
			$E_j^0 ~^*$ & 2.46,~~2.38,~~2.30,~~2.15,~~1.98 & $V$ \\
			&$i_j^0~^{**}$ & 2.00,~~0.02,~~0.02,~~0.02,~~0.02 & $A/m^2$ \\ \hline
			\multirow{9}{*}{\textbf{All Models}} &  
  			$v~^*$ & 0.0114 & $L$ \\
			& $S_{sat}~^*$ & 0.0001 & $g$ \\
			& $a_v^0~^*$ & 1 & $m^2$ \\
			& $\gamma~^*$ & 1.5  & - \\
			& $\omega~^{**}$ & 0.1 & $1/g$ \\
			& $k_p~^*$ & 22 & $1/(g s)$ \\
			& $R$ & 8.3145 & $J/(K~mol)$ \\ 
			& $F$ & 9.649$\times10^4$ & $C/mol$ \\ 
			& $T$ & 298 & $K$ \\ 
			\hline
		\end{tabular}
	\begin{tablenotes}
      \item $^*$ adopted from \cite{zhang2015modeling,marinescu2016zero}, $^{**}$ estimated parameters.
    \end{tablenotes}
  \end{threeparttable}
\end{table}

The main observations from simulation results are as follows. First, all the reaction models provide similar specific capacity values, approximately 1675 $mAh/g$, matching the theoretical specific capacity value in \cite{rezan2017li}. Second, the intermediate reactions associated with the high plateau can influence this plateau's shape and duration. Specifically, when the high plateau reaction is simplified to only $S_8 \rightarrow S_4^{2-}$ in Model \#1, the voltage curve shows a relatively longer duration for this plateau. Moreover, this high plateau ends with a dramatic, almost vertical voltage drop. The same observation is presented in \cite{hofmann2014mechanistic}. This drop can be explained by the potential dynamics given by Eqn. \ref{dotEj}. At the end of the reaction, the mass of the reactant approaches zero, while the rate of change of this mass maintains a relatively stable value due to the fact that $i_j$ still needs to meet the external discharge current value. This results in $\dot{E}_j \rightarrow -\infty$. Hence, the measured voltage $V$ exhibits an almost vertical drop. When intermediate reactions are added between $S_8$ and $S_4^{2-}$, a milder slope emerges with the dip between the plateaus occurring at higher voltages.
\begin{equation}
    \dot{E}_j =  - \frac{RT M_s v}{n_j F} \sum_i s_{i,j} n_{S_i} \frac{\dot{m_i}}{m_i}  \label{dotEj}
\end{equation}

This analysis provides the characteristics of the voltage curve considering different reactions. The bottom line is that simplified representations of the reduction reactions are reasonable in terms of one's ability to capture theoretical specific capacity, but the influence of different simplifications on the voltage features is significant. One needs to choose the reactions carefully based on the purpose of the given application. For example, studies focused on the capacity fade may choose simpler models, while state of charge estimation may require more complicated reaction models due to the accuracy required to capture the characteristics of output voltage. 

\section{Parameter Sensitivity Analysis}

One can broadly categorize the parameters of the above models into two groups. The first group includes parameters that are known \textit{a priori}, such as Faraday's constant, gas constant, temperature, molar mass, number of electrons transferred in each reaction, etc. The second group contains unknowns such as standard potentials, exchange current densities, initial masses of the dissolved species, and precipitation-relevant parameters (i.e., the morphology parameters $\gamma$, $\omega$, and precipitation rate). Increasing the number of reactions and species considered in the model increases the number of unknown parameters. Hence, the difficulty of parameterization increases. In this section, the authors analyze the sensitivity of the 0D model's output to the underlying parameters in order to better understand which parameters are critical to identify.

Instead of using a theoretical parameter sensitivity approach \cite{doosthosseini2019accuracy}, this work applies a similar analysis method to the work presented in  \cite{ghaznavi2014sensitivity,ghaznavi2015sensitivity,ghaznavi2014analysis} for a 1D Li-S battery model. Specifically, the authors choose one parameter at a time and investigate the voltage performance changes due to variations in only this parameter. All other parameters are held constant. This parameter sensitivity analysis is performed on all models listed in Table \ref{tab:Reactions}. Since Model \#4 describes the full reaction chain containing the largest number of parameters (total number of 14), the results of Model \#4 are chosen for the following sensitivity discussions. Cell voltage sensitivities to different parameters are plotted for two perturbations from the nominal values.  

\subsection{Sensitivity to Standard Potential $E_j^0$}
\begin{figure*}[ht]
    \centering
    \includegraphics[width=1\textwidth]{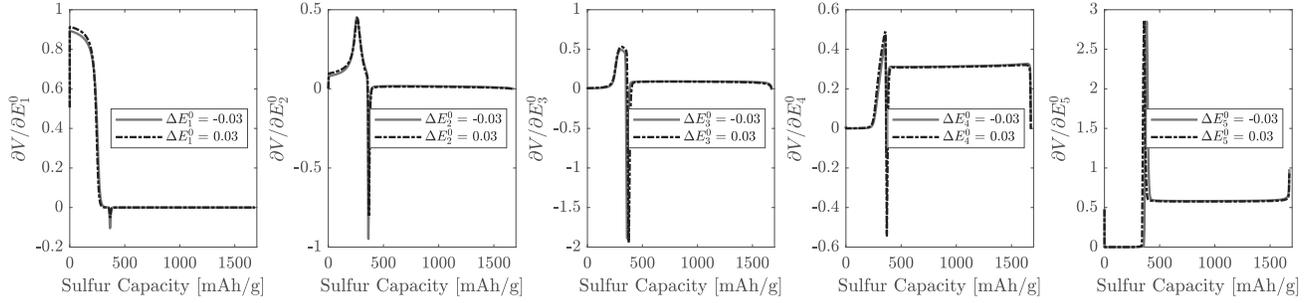}
    \caption{Cell voltage changes with respect to the deviations of nominal standard potential $E_j^0$ for Model \#4}
    \label{fig:Ej_model4}
\end{figure*}

Fig. \ref{fig:Ej_model4} presents the results of analyzing the cell voltage sensitivity of Model \#4 to parameter $E_j^0$. One can see the relative dominance of different reactions by observing the corresponding changes in the voltage curve. Reaction $S_8 \rightarrow S_8^{2-}$ ($j=1$) has a dominant influence on the high plateau voltage level. Reaction $S_8^{2-} \rightarrow S_4^{2-}$ ($j=2,3$) has a strong effect on the ``ending slope" of the high plateau and the dip point between the two plateaus, while slightly changing the low plateau voltage level. The last two reactions ($j=4,5$) occur during the low plateau region, and therefore both of the corresponding standard potentials influence the voltage curve dramatically. 

\subsection{Sensitivity to Exchange Current Density $i_j^0$}
\begin{figure*}[ht]
    \centering
    \includegraphics[width=1\textwidth]{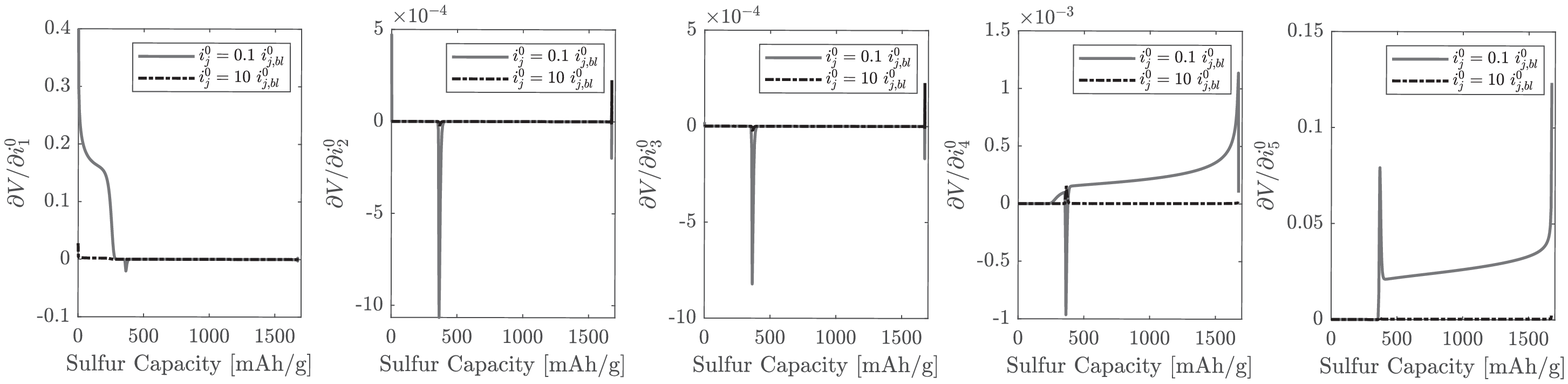}
    \caption{Cell voltage changes with respect to the deviations of nominal exchange current density $i_j^0$ for Model \#4}
    \label{fig:ij_model4}
\end{figure*}

The sensitivity analysis for $i_j^0$ is shown in Fig. \ref{fig:ij_model4}. One can observe minor changes in the voltage curve within the chosen range of $i_j^0$. The exchange current densities of the middle reactions have negligible impacts on voltage, as the voltage curves show a negligible difference. Compared to $E_j^0$, which influences the voltage $V$ through the variable $E_j$, the parameter $i_j^0$ affects the overpotential $\eta_j$ though an exponential function. On the one hand, this shows that $E_j^0$ and $i_j^0$ have similar effects on voltage, and highlights the challenge of identifying both $i_j^0$ and $E_j^0$ simultaneously from experimental data. On the other hand, bigger variations in $i_j^0$ produce smaller changes in the voltage curve compared to $E_j^0$, which suggests the exchange current density has less of an impact on the model's voltage performance.

\subsection{Sensitivity to Precipitation-related Parameters}
\begin{figure*}[ht]
    \centering
    \includegraphics[width=0.9\textwidth]{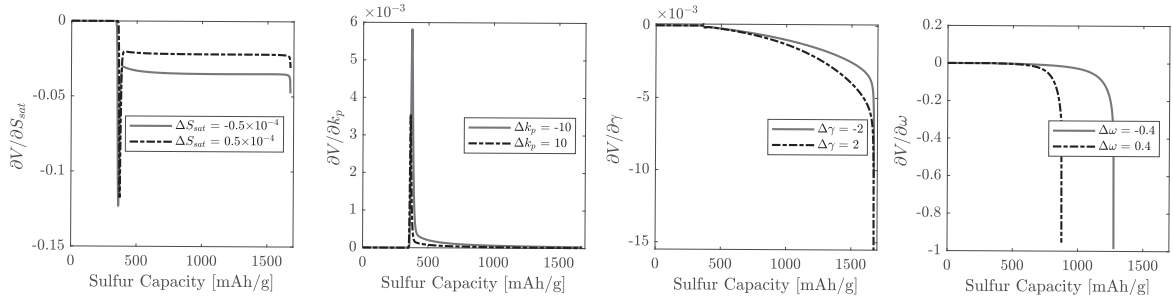}
    \caption{Cell voltage changes with respect to the deviations of nominal precipitation relevant parameters for Model \#4}
    \label{fig:pre_model4}
\end{figure*}

Four parameters essentially govern the precipitation phenomenon, namely, the saturation mass of dissolved $S^{2-}$, the precipitation rate $k_p$, and the morphology parameters of the relative porosity of the cathode, $\gamma$ and $\omega$. The variations of each parameter produce changes in the low plateau region without influencing high plateau voltages, as shown in Fig. \ref{fig:pre_model4}. When the saturation mass increases, more $S^{2-}$ is dissolved in the electrolyte, reducing the potential of the reaction $ S_x^{2-} \rightarrow S^{2-} $, hence providing a lower cell voltage. The precipitation rate $k_p$ and morphology parameter $\gamma$ determine the flatness of both the beginning and end regions of the low plateau. Larger values of $k_p$ and smaller values of $\gamma$ provide a flatter low plateau. The parameter $\omega$ is the only parameter dramatically affecting the capacity of $S_8$. With an increase in the rate of change of relative porosity, the discharge process can terminate earlier because of the resulting paucity of active reaction area inside the cathode material.

\subsection{Sensitivity to Initial Mass of Dissolved $S_8$}
\begin{figure}[ht]
    \centering
    \includegraphics[width=0.4\textwidth]{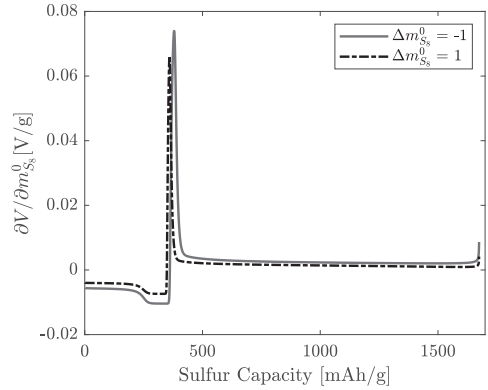}
    \caption{Cell voltage changes with respect to the deviations of nominal initial mass of dissolved $S_8$ for Model \#4}
    \label{fig:ms_model4}
\end{figure}

The authors assume that all the active sulfur is dissolved into the electrolyte at the beginning of discharge. The mass of the dissolved $S_8$ is the main factor determining the total amount of active material inside the battery, and the initial masses of other species at the beginning of discharge are comparatively very small. In this case, sensitivity analysis is only performed with respect to the initial mass of dissolved $S_8$. The results are presented in Fig. \ref{fig:ms_model4} with negligible impacts on the voltage curve and specific capacity. However, one must note that the voltage curve difference emerges when plotting the voltage with respect to time instead of specific sulfur capacity. Hence the dip point between the two plateaus changes with respect to time.

From this analysis, the following conclusions are made to guide parameter identification: (i) Although both $E_j^0$ and $i_j^0$ affect the voltage curve, $V$ is more sensitive to $E_j^0$. (ii) In the set of parameters relevant to precipitation, correctly capturing the values $\omega$ and $\gamma$ is more essential for capturing the shape of the low plateau. (iii) The initial mass of dissolved $S_8$ influences the timing of the dip point between the high and low plateaus. Hence, in the parameter identification study, the authors focus on parameters $E_j^0$, $\gamma$, $\omega$, and $m_{S_8}^0$, with all the other parameters treated as known constants.

\section{Parameter Identification Study}

It is difficult to systematically compare multiple models of Li-S battery dynamics without fitting all of them to the underlying experimental data first. This section presents a parameterization study focusing on the proposed four reaction models. The authors obtain the test data through discharging laboratory-made prototype coin cells.

\subsection{Coin Cell Fabrication and Discharging Experiments}
The authors fabricated 2016-type coin cells using the recipe in \cite{li2017organosulfide}. The cathode material consisted of Carbon/Sulfur composite with 70 wt\% sulfur, Ketjenblack EC60JD to form a conductive framework, and polyvinylidene fluoride (PVDF) dissolved in N-Methyl-2-pyrrolidone (NMP) as a binder. A lithium chip of 0.6 mm thickness served as the anode. The electrolyte used here was 1M $LiTFSI$ and 4 wt\% $LiNO_3$ in the dioxolane/dimethoxylethane mixture (DOL/DME = 1:1, V/V). The sulfur loading of the cell was calculated at 0.521 mg. A constant discharge current of 0.03 mA was applied in the experiment to obtain the discharge voltage curve. Testing was conducted at room temperature, using ARBIN battery test equipment with special attention given to the accuracy of current control. In particular, a separate capacitor charge/discharge test was used for estimating the bias in the cycler's current measurements, and this bias was used to correct measured data. 

\subsection{Scaling From Model to Prototype Using Similitude}
One caveat appears when we implement the experimentally-measured Li-S cell current into the simulation model, namely, the need to scale the simulation model to match the physical sizing of the Li-S coin cell used in this study. Here we use the idea of similitude to scale the relative variables/parameters from the model to the prototype. For the fundamental dimensions in the model (i.e. time, mass, charge, length and temperature), charge is scaled by a given value $\mu$, and we want to find the scaling factor for mass and length, with the other dimensions unchanged. Then the input current is scaled by $\mu$ due to the fact that time is not scaled. The relationship between the current inputs of the model and prototype is shown in Eqn. \ref{current_scale} and applies to $i_j$ as well. The subscripts $mod$ and $pro$ represent model and prototype correspondingly. 
\begin{equation}
    I_{mod} = \mu~I_{pro} \label{current_scale}
\end{equation}
From Eqn. \ref{mass1}, we derive the scaling law for mass as shown below:
\begin{equation}
   \frac{\dot{m}_{i,mod}}{\sum_j \frac{n_{S_i}}{n_j}s_{i,j}~i_{j,mod}} = \frac{\dot{m}_{i,pro}}{\sum_j \frac{n_{S_i}}{n_j}s_{i,j}~i_{j,pro}}
\end{equation}
where other parameters are independent of mass and charge, except for $\dot{m_i}$ and $i_j$. Hence, we have
\begin{equation}
    m_{mod} = \mu ~m_{pro} 
\end{equation}
Using the same logic, one can derive the scale factor for length as $\mu^{\frac{1}{3}}$ from Eqn. \ref{ij}, and the scale factors for parameters $a_v$, $v$, $i_j^0$, $\omega$, $k_p$ are $\mu^{\frac{2}{3}}$, $\mu$, $\mu^{\frac{1}{3}}$, $\frac{1}{\mu}$ and $\frac{1}{\mu}$. In the following parameterization simulation, the authors use 1A as the simulated model's input current with the corresponding $\mu$ being $3.33\times 10^{4}$. This value of $\mu$ scales the simulation model down to match the true scale of the fabricated laboratory coin cell. 

\subsection{Optimization Problem Formulation}

The parameter identification study uses the least-squares method based on the above 0D reaction models. The least-squares method estimates parameters by minimizing the squared discrepancies between observed data and the expected values. Given the models' highly nonlinear structure, the authors choose the particle swarm optimization algorithm \cite{kennedy1995particle} to solve the resulting nonlinear least-squares problem. The goal is to identify the parameter vector $\bm{\hat{\theta}}$ by solving the problem below:

\begin{equation}
\min_{\bm{\hat{\theta}}} J= \sum_{k=1}^{N_{min}} (\hat{V}_k-V_{m,k})^2 + \alpha (\hat{T}-T)^2
\label{LSE}
\end{equation}
\begin{equation}
where: \bm{\theta} = [~E_j^0,~\gamma, ~\omega, ~m_{S_8}^0~]^T, ~~N_{min} = min \{N, \hat{N}\}
\end{equation}
subject to the dynamics of the system from Eqn. \ref{mass1} to \ref{VoltageEqn}. In this problem formulation, $\hat{V}_k$ and $V_{m,k}$ represent the estimated and measured output voltages at time step $k$; $\hat{T}$ and $T$ are the final times of the simulation and experiment; and the weight $\alpha$ reflects the importance of minimizing the difference between the simulated and experimental discharge durations. 

\subsection{Parameter Identification Results}

\begin{figure*}[ht]
    \centering
    \includegraphics[width=1\textwidth]{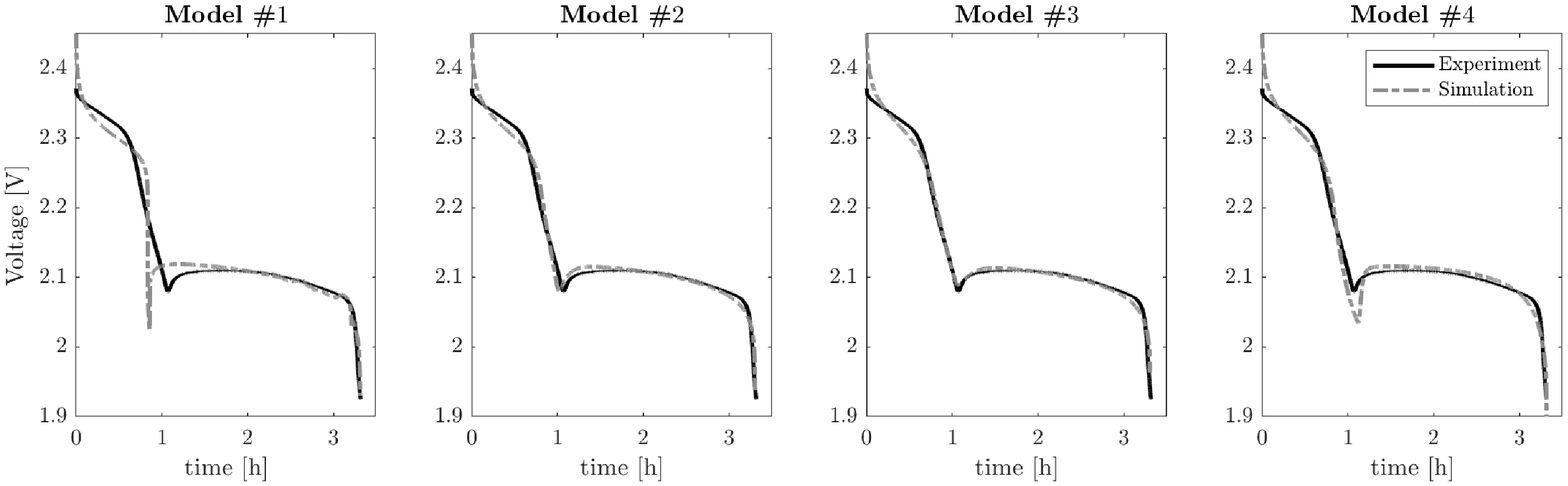}
    \caption{Discharge voltage comparison between experiment and simulation using identified parameters for all the models}
    \label{fig:ParaID_voltage}
\end{figure*}

\begin{figure}[ht]
    \centering
    \includegraphics[ width=0.8\textwidth]{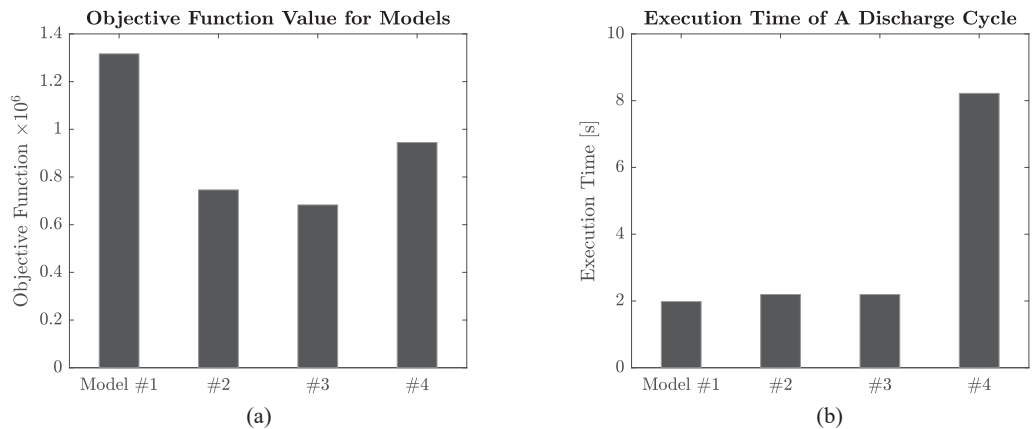}
    \caption{(a) Objective function values for the models~~~~(b) Execution time of one discharge cycle for the models}
    \label{fig:CostTime}
\end{figure}

The identification results are listed in Table \ref{tab:para_id}, with the corresponding estimated voltage shown in Fig. \ref{fig:ParaID_voltage}. The identified parameters for all the models are able to fit the experimental data in terms of the total discharging time. The objective function values corresponding to the best estimated parameters for the models are shown in Fig. \ref{fig:CostTime} (a). Interestingly, Model \#3 performs the best fit with a minimum objective function value among all the models. The over-simplification of reactions in Model \#1 leads to failure in capturing the slope of the voltage in the high plateau region. In contrast, the full reaction Model \#4 does not provide a good fit in the middle region where the precipitation phenomenon emerges. This is potentially due to the simplification of the precipitation dynamics in the model. A more sophisticated physics-based precipitation model may improve the accuracy of the full reaction model.

\begin{table}
	\begin{center}
		\caption{Parameter identification results}
		\label{tab:para_id}
		\begin{tabular}{|C{1.5cm}|C{1.6cm}|C{2.6cm}|C{3.5cm}|C{4.5cm}|C{0.7cm}|}
			\hline 
			\textbf{Parameter} &\textbf{Model \#1} &\textbf{Model \#2} &\textbf{Model \#3} &\textbf{Model \#4} & \textbf{Units} \\
			\hline
			$E_j^0$ & 2.453,~2.090 & 2.464,~2.360,~2.073& 2.467,~2.374,~2.342,~2.069& 2.467,~2.310,~2.208,~2.094,~1.972& $V$ \\ 
			$\gamma$ & 0.316 & 0.437 & 0.483 & 0.959 & -\\ 
			$\omega$ & 0.508 & 0.570 & 0.613 & 0.650 & $1/g$ \\ 
			$m_{S_8,mod}^0$ & 2.001 & 2.628 & 3.038 & 3.523 & $g$ \\ 
			$m_{S_8,pro}^0$  & 0.060 & 0.079  & 0.091 & 0.106 & $mg$ \\
			\hline
		\end{tabular}
	\end{center}
\end{table}

Another observation pertains to the simulation time needed for one discharge cycle, as shown in Fig. \ref{fig:CostTime} (b). Intuitively, the computational complexity of the models increases with the number of included reduction reactions. However, the execution time increases from Model \#1 to Model \#3 are minor, whereas the increase from Model \#3 to Model \#4 is quite significant. One can conclude the best fit for further control application is Model \#3, as it provides the least discrepancy between the experimental data and simulation results, without excessive computational cost. 

\section{Conclusions}
This paper performs parameter sensitivity analysis and identification for a variety of zero-dimensional physics-based Li-S battery models. The authors provide answers to the questions listed in Introduction: (i) The intermediate reactions between $S_8$ and $S_4^{2-}$ add slope to the second half of the high plateau, while  the reactions between $S_4^{2-}$ and $S^{2-}$ prolong the flat portion of this plateau and deepen the inter-plateau voltage dip. (ii) Sensitivity analysis provides guidance on the final choice of to-be-identified parameters as $E_j^0$ (the voltage level), $\gamma$ (the low plateau slope), $\omega$ (the low plateau duration) and $m_{S_8}^0$ (the dip location). (iii) Taking into account the intermediate reactions affecting the high plateau, the estimated voltage curve fits the experimental data well except at the beginning of discharge. Finally, taking into account the importance of balancing parameter identification performance versus execution time, the authors recommend Model \#3 as the best model for use in future control oriented applications under this model structure.

\begin{acknowledgment}
This work is funded by National Science Foundation Grant 1351146. The authors gratefully acknowledge this support. Any opinions, findings, and conclusions or recommendations expressed in this material are those of the authors and do not necessarily reflect the views of the National Science Foundation.
\end{acknowledgment}

\bibliographystyle{asmems4}
\bibliography{reference}



\end{document}